# AN ANALYSIS OF A BERT DEEP LEARNING STRATEGY ON A TECHNOLOGY ASSISTED REVIEW TASK

Alexandros Ioannidis[1]

[1]Computer and Information Sciences Department, University of Strathclyde, Glasgow, UK
alexandros.ioannidis@strath.ac.uk

## ABSTRACT

*Document screening is a central task within EBM (Evidence-based Medicine), which is a clinical discipline that supplements scientific proof to back medical decisions. Given the recent advances in DL (Deep Learning) methods applied to IR (Information Retrieval) tasks, I propose a DL document classification approach with BERT (Bidirectional Encoder Representations from Transformers) or PubMedBERT embeddings and a DL similarity search path using SBERT (Sentence-BERT) embeddings to reduce physicians' tasks of screening and classifying immense amounts of documents to answer clinical queries. I test and evaluate the retrieval effectiveness of my DL strategy on the 2017 and 2018 CLEF eHealth collections. I find that the proposed DL strategy works, I compare it to the recently successful BM25+RM3 (IR) model, and conclude that the suggested method accomplishes advanced retrieval performance in the initial ranking of the articles with the aforementioned datasets, for the CLEF eHealth Technologically Assisted Reviews in Empirical Medicine Task.*

## 1. INTRODUCTION

Nowadays, the communication of analytically substantial research is restricted by the geometric development of the medical information. EBM aims to provide a solution by integrating the experience of the clinician and the best available scientific information to guide decision-making. While EBM is a keystone element of medical practice, medical SRs (Systematic Reviews) are the corner stone of EBM. The Cochrane Library defined SR as an effort to establish the identity and worth of all observed facts also known as empirical evidence. According to Tsertsvadze et al., new research studies provide alternative methods to minimise the time and cost associated with all SR (Systematic Review) stages [1].

TAR (Technologically Assisted Reviews) is a category of SRs which could give a solution to the above-mentioned research restriction. This could happen by efficiently identifying, selecting, assessing, accumulating and examining relevant research evidence which is admitted in the analysis. According to Wallace et al. [2], the formation of SRs without considering the fast accumulation of research documentation is evidently an arid exercise. For example, the average time to complete a SR is between 2 to 6 months and the cost of screening SRs is estimated in tens and hundreds of thousands of pounds as it requires shifting through thousands or tens of thousands of medical articles based on Tsertsvadze et al. [1]. The TAR objective is to minimise the overall cost involved in assessing articles, by finding all the relevant pieces as early as possible and to identify the relevant items as accurately as possible with the minimum effort.

An additional benefit of improving the screening effectiveness using TAR is the minimisation of publication bias in the SR findings report. According to Hardi and Fowler this is accompanied by several positive aspects such as enabling the assembly of useful results to assist researchers in the strictness, breadth and depth of their literature reviews and provide a critical methodology for the advancement of EBM healthcare [3]. Current approaches to solve SR challenges are based on ranking the documents, classifying the documents, or using AL (Active Learning) based approaches. However, with advances being made in DL and the recent and consecutive successes of BERT on several NLP (Natural Language Processing) and IR tracks, I propose a BERT based DL initial ranking strategy specifically for the CLEF eHealth Task 2. CLEF eHealth Task 2 is the CLEF eHealth Technologically Assisted Reviews in Empirical Medicine Task.

My research addresses key challenges in SRs by supporting reductions to the amount of process time and cost factors to perform a TAR. Specifically, the current investigation supplements tested proposals to enhance the initial screening stage of biomedical documents (e.g. PubMed data) to achieve improved screening effectiveness with less human effort when conducting SRs. My approach is described, evaluated and compared against an effectively customised BM25+RM3 (Relevance Model 3) approach.

My suggested IR method promotes the combined usage of a fundamentally different text representation (distributed) and a pretrained DL model (BERT, PubMedBERT) on PubMed abstract data to form an alternative strategy for the CLEF eHealth Task 2 topics. The SBERT model, which is an extension of the BERT model, was also investigated as it has proven to be a promising model due to its effectiveness on text processing, NLP and IR sentiment analysis tasks with the MS MARCO and MultiNLI datasets.

For the preceding reasons, I explored a DL strategy awaiting a new SOTA method for the initial ranking of the CLEF eHealth TAR Task 2. The hypothesis was that using a DL method on biomedical SRs will result better evaluated CLEF-TAR metrics than the BM25+RM3 method. If my DL approach improves the retrieval efficiency (CLEF-TAR accuracy) and screening workload, it could lead potentially to further reduction of cost and time required for SRs assessments. Therefore, the following research question was raised and researched:

How does a DL strategy with BERT influences the retrieval performance at the initial screening stage in a TAR task?

## 1.1. A description of the CLEF eHealth for TAR in Empirical Medicine Task

This IR track focuses on abstract and title screening and has two objectives. First, to produce an efficient sorting of the articles, such that all the relevant abstracts are retrieved as quickly as possible. Second, to identify a subset of the A set which contains all or as many of the relevant abstracts for the least effort. The A set is the result of potentially relevant studies when a CLEF query is submitted to the PubMed database containing titles and abstracts of medical studies. Furthermore, the participants are given the associated judgements file (QRELS), which has the following TREC format: Topic ID, Iteration, PubMed Document ID (PMID), Relevancy and a Binary code of 0 for not relevant documents and 1 for relevant documents. The order of PMIDs in the assessments file does not suggest degree of relevance. Lastly, documents not occurring in the QRELS file are not judged by the human assessor and therefore are presumed to be irrelevant in the evaluations.

## 1.2. Organizational Structure

In this part, I provide a brief summary of the remaining sections of the article. Section 2 provides an overview of the previous methods for the CLEF eHealth in empirical medicine task for example different indexing, query formulation and retrieval processes. Furthermore, section 2 describes previous DL approaches for IR tasks. Section 3 outlines the general methodology used including the data and materials, my baseline retrieval model, the DL method I used for document retrieval, the measures utilised and the statistical comparison process I followed. Section 4 demonstrates the statistically significant results from my evaluation of the proposed DL strategy. Finally, section 5 is the concluding piece of the article with a summary of the work that was undertaken, the results obtained and the potential paths for future work.

## 2. RELATED WORK

### 2.1. Previous CLEF eHealth Strategies

Alharbi and Stevenson utilised the review title and the Boolean query to order the abstracts retrieved with the query by applying standard similarity measures. According to Alharbi and Stevenson, the title and terms extracted from the Boolean query contributed the most useful information for this task [4]. Their methodology used three topic parts, the title text, the terms and Mesh items parsed from the Boolean query. The Mesh items were pre-processed with the same approach applied on the Boolean query. Moreover, pre-processing was added to the PubMed abstracts and data extracted from the topics. The text was tokenised, changed to lower case, stop words were taken out and the rest of the tokens were stemmed. Lastly, the extracted data from the topic and all abstracts was converted into TF-IDF weighted vectors to calculate similarity among the topic and every abstract using the cosine metric for the pair of vectors [4].

Chen et al. used a customised L2R (Learning-to-Rank) model and the Word2vec NLP technique to represent queries and documents in order to compute their similarities by cosine distance [5]. Their L2R method consisted of three points: i) query expansion, ii) feature extraction and iii) model training. Later, in the query expansion part, the authors improved retrieval precision by expanding queries. In the feature extraction stage, they extracted features of each query document pair. When a document was retrieved under a query, it was connected with a weighting score and rank. Finally, in the learning phase of the L2R model, the relevance of a query-document pair was assessed with a random forest classifier. An alternative approach was proposed by Singh et al. [6] using QE (Query Expansion) based on relevance feedback and TF-IDF similarity score.

Norman et al. used a system that built on logistic regression, and implemented strategies to handle class imbalance and perform relevance feedback [7]. Additionally, they tested two classifiers, logistic regression with stochastic gradient descent learning on the entire data and standard logistic regression trained using conventional methods on a sub group of the data with added pre-processing to enhance the yield. Nunzio et al. concentrated their work [8] on discovering the optimal combination of hyper-parameters by utilising the training data available and a brute force strategy. Moreover, this strategy created different query features of the identical information need given a minimum amount of relevance feedback. Furthermore, the authors of the paper [8] decided to use a plain BM25 retrieval model and acquire the relevance feedback for the first abstract in the ranking list for each topic. They built two different queries according to the value of the feedback and combined these with the original one in different ways. Lastly, they designed alternative strategies which included as parameters the number of documents to assess in batches, the percent of documents to assess, the maximum number of documents to assess per iteration, the number of terms to add at each feedback iteration for the cost-effective evaluation and the minimum precision the system can reach before stopping the search.

Additionally, Singh et al. [6] utilised Lucene's inverted index to index the retrieved articles from the PubMed query search while performing a SR. The query was processed for term boosting, fuzzy search and used for scoring documents according to TF-IDF similarity. Relevance feedback is used to update the query and become more pragmatic. Moreover, an SDR (Seed-Driven Document) ranking model for screening prioritisation was proposed by Lee et al. [9] with the assumption that one relevant document is known. This SDR model differentiated itself from other solutions by representing data using a BoCW (Bag of Clinical Words) method instead of the traditional bag of words approach. In the same research paper [9] the authors reported to achieve the best ranking when they combined SDR with word embedding relevance. The authors concluded that each seed document they used showed different ranking performance and they pointed out a future direction on how to incorporate additional usable resources to enhance the seed document.

**2.2. Deep Learning for Information Retrieval Tasks**

A different approach using CNN (Convolutional Neural Networks) for PubMed article classification has been introduced by Lee and Eunkyung [10]. This NN (Neural Network) approach enhanced the performance of identifying relevant documents for SRs. The three models proposed in this study [10] incorporated the concatenated extracted context from criteria and clinical documents. The authors inferred from the evaluated results that while the performance of the suggested approaches had room for improvement, they managed higher performance in full document screening than abstract screening. This work [10] in particular motivated my research because it was a step towards applying DNNs (Deep Neural Networks) to improve the SR screening stage in spite of the scarcity of labelled documents and the imbalanced data in the CLEF collection which has also been pointed out by Cormack and Grossman [11].

Further research has been conducted by Carvallo et al. [12] to compare the performance of alternative language models based on two neural word embeddings (Word2vec and GloVe) for document representation in an AL environment. They evaluated these methods with CLEF eHealth 2017 and Epistemonikos datasets. They concluded that the Word2vec approach had less variance and better general performance than GloVe when using AL strategies with uncertainty sampling on the biomedical document screening task [12]. Carvallo et al. continued to investigate other AL strategies incorporating BERT and BioBERT to evaluate the results with the TAR toolkit [13]. Later, the authors inferred that neural embeddings in an AL setting outperform the traditional TF-IDF representation [13]. When compared among neural and textual embeddings with the CLEF eHealth 2017 dataset, BERT and BioBERT yielded the best results in an AL setup. They also inferred that an uncertainty sampling strategy combined with logistic regression achieved the best performance overall [13]. Finally, their comparison results indicated that in terms of last relevant document their method US-RF-BioBERT was the best. However, for different CLEF-TAR metrics such as the AP (Average Precision) and the NCG (Normalised Cumulative Gain) after certain documents, the authors reported that their method did not perform well [13]. Essentially, different people have implemented different strategies and they all used different methods, so itis hard to determine which IR strategy is the absolute best to use for the CLEF eHealth Task 2.

Nonetheless, pretrained DL architectures incorporating BERT proposed by Devlin et al. [14] have advanced performance in several NLP tasks with the MS MARCO and MultiNLI datasets. In fact, pretrained neural language models like BERT are recognised as the basis of SOTA NLP methods [15]. Specifically, BERT in a default set up has managed to exceed existent retrieval algorithms with the Robust04 dataset, on tasks that have not seen IR effectiveness advances over classic IR baselines for a long-time in accordance to Camara and Hauff [16]. BERT uses the Transformer mechanism which has a necessary encoder for reading text input to generate a LM (Language Model) and an optional decoder for producing task prediction. Furthermore, BERT has been investigated for some NLP and IR tasks in the context of Evidence retrieval and claim verification by Soleimani et al. [17], Rethinking query expansion for re-ranking by Padaki et al. [21], Retrieval heuristics by Camara et al. [16], Augmenting it with Graph Embedding (VGCN-BERT) for Text Classification by Lu et al. [18] and Alleviating Legal News Monitoring by Sanchez et al. [19].

Also, Akkalyoncu et al. introduced the Birch framework [20] a system that integrates neural networks (BERT) to document retrieval using the open-source Anserini [21] toolkit for IR to present end-to-end search over large document collections. Birch adopts an architecture that incorporates the Anserini IR toolkit for initial retrieval, and then inference using an applied BERT model. Birch implements simple ranking models which have achieved SOTA effectiveness on standard TREC newswire and social media test collections. According to Yang et al. [22] this framework has been a highly successful IR toolbox particularly with the Microblog and Robust04 collections supporting simple applications of BERT for ad-hoc document retrieval [22].

Additionally, Roy et al. [23] experimented with FastText which learns the embedded representations of character n-grams before integrating these to attain word vectors and contradicted their model against the aforementioned Word2vec. The authors of the study [23] noticed that the likelihood of receiving improved IR effectiveness with Word2vec embeddings from normalized target collections did not apply in the case of FastText. Further, they observed that retrieval performance was generally better with Word2vec on stemmed collections (e.g. TREC-Rb and WT10G). When in fact, FastText returned better retrieval performance on unprocessed collections (e.g. Gov). Also, Palangi et al. in their study [24] proposed a recurrent neural networks model with LSTM (Long Short-Term Memory) cells. The LSTM-RNN model was suggested by the authors due to its ability to maintain long term memory while accumulating more information while processing the sentence until the last word. Thus providing a semantic representation of the complete sentence. Palangi et al. proved that their strategy on the web document retrieval task, outperforms all existing state of the art methods significantly.

Furthermore, Reimers and Gurevych introduced the open source SBERT [25] model (part of the sentence-transformers scheme) which is a modification of the pretrained BERT network that uses siamese and triplet network structures to derive semantically meaningful sentence embedding that can be compared using cosine-similarity. Contrary to traditional search engines which only find documents based on lexical matches, semantic search with SBERT can also find synonyms. Likewise, to BERT, the SBERT scheme has been evaluated on common transfer learning tasks, where it outperformed other SOTA sentence embeddings methods [25]. Additionally, SBERT has been reported to reduce the effort for finding the most similar pair from 65 hours with BERT to about 5 seconds while maintaining the accuracy from BERT. Therefore, making it a considerably more efficient networks transformers scheme than other specified document retrieval solutions in terms of execution time [25]. Thus, I was prompted to include SBERT in my experiments due to the time-wise practicality and its promising accuracy. From my study of the literature, I understood that BERT based methods have proven to be a promising approach to document classification based on previous IR task experiments with different domain datasets.

## 3. METHOD

### 3.1. Data and Materials

I used the CLEF eHealth 2017 and 2018 collections (www.clef-initiative.eu/dataset/test-collection, [26]) for the TAR eHealth Task 2. The collections are composed of approximately 350,000 PubMed abstract articles which were constructed based on 100 systematic reviews (each systematic review is a topic). The average number of documents per topic was about 4,000 documents. The CLEF 2017 TAR Task 2 introduced 50 topics [27] and the analogous CLEF 2018 TAR Task presented 50 additional topics [28].

Related to Indexing: Each PubMed document is composed of a number of fields. For this study, the following fields were indexed Title, Abstract, PMID, Author(s), Journal Title, Year, Mesh Heading List (MHL) and MedlineTA fields of the PubMed SRs. I tokenised the text, changed it to lower case, removed stop words and I stemmed the remaining tokens. I also used the 2017 NYT Common Core collection for experiment validation purposes which was processed in a similar fashion. Lastly, the data extracted from all articles and topics were converted into TF-IDF weighted vectors [9] to calculate the similarity among the topic and every document using the cosine metric for the pair of vectors.

With respect to Topics and Queries: Every topic of the CLEF collection contained the topic-ID, the review title, the Boolean query manually drafted and constructed by Cochrane experts (www.training.cochrane.org/ask-expert, [29]) and the collection of PMIDs (PubMed Document Identifiers) retrieved by executing the query in MEDLINE databases. For this study, I created queries for each topic by concatenating the title, and pre-processed Boolean query together and then expanded this query. For the pre-processing of the Boolean queries, I stripped out all symbols e.g. +, *, etc. (www.nlm.nih.gov/bsd/mms/medlineelements.html, [30]), and only kept the terms, while abbreviations were expanded. I found that this pre-processing resulted in improved performance.

To further improve the initial query, I further expanded the query according to the Thesaurus based QE method suggested by Imran et al. [31]. This method has been used in many IR systems for identifying linguistic entities that are semantically close and has increased the effectiveness of the retrieval performance in IR tasks [31]. Thus, I focused on improving the queries that I pass on the tested search engine system by applying a Thesaurus based QE. I used TF-IDF vector scores to rank the document collection terms for each CLEF topic to simulate my Thesaurus based QE technique. Afterwards, I selected a fixed size of the highest scoring terms and added these to the concatenated title and pre-processed query for each CLEF topic.

In relation to Tools: My research included experimentation with document indexing, retrieving and ranking, query parsing, pre-processing and expansion methods. In addition, I replicated a baseline and extended open source toolkits to demonstrate the proposed DL method for the examined task. I considered variances of my suggested DL proposal against a currently advanced method in the investigated field of initial ranking for the CLEF eHealth TAR in Empirical Medicine task.

### 3.2. Retrieval Model

The baseline model implementation for this paper is the BM25 plus query likelihood with RM3 expansion [32]. The exact RM3 parameter settings I used can be accessed through the link in the reference [33]. I set the BM25+RM3 model's retrieval parameters with the default Anserini values (k1=0.9, b=0.4 for BM25 and fbTerms=10, Weight=0.5 for RM3) and I compared this to my DL approaches.

### 3.3. Deep Learning for Document Retrieval

In summary, I experimented with the following approaches:

1. Document Classification with Birch. Classifying each document (actually each sentence in the document), predicting its relevance, and combining the classifier scores with the original relevance score. The classifier takes the previously formulated query (described in subsection 3.1) plus a seed document as input, to predict the ranking score.

    a. I used the Bert-Base uncased word embeddings which have been trained with 12 layers, 768 hidden, 12 heads and 110M parameters as described by Devlin et al. [14]. The maximum sequence length was set to the default value. The optimization method I used was BertAdam with warmup steps before the learning rate scheduler. The learning rate was equal to 1e-5. A good description of the DL architecture has been provided by Akkalyoncu et al. [20]. Additionally, I set the batch size during training and inference to 8 and 4 documents accordingly. I used 3, 6 and 12 evaluation steps with one epoch. Lastly, the train loss I noticed after having finished training was estimated to be 0.8291 on average.

b. Overall, I explored the Birch toolkit two language models using my two CLEF datasets. The language models I used were the aforementioned vanilla BERT model and afterwards the PubMedBERT model which is described in the end of this subsection (3.3.3.).
2. Similarity Search score with sentence-transformers for each document returned. The similarity based SBERT [25] model took two inputs the sentence of a document and the query. Then, it computed the similarity score between the embeddings to get a score, and sum over all the scores. Finally, I combined this score with the original relevance score (BM25+RM3 scores).

### 3.3.1. Using Birch with the CLEF dataset

I extended the Birch toolbox to make it work with CLEF collections and topics and then evaluated my TREC formatted results with the CLEF-TAR toolkit. For the initial retrieval, I used my baseline input (initial ranking from BM25 plus query likelihood with RM3 relevance feedback [32]), which provided the initial ranking to depth 1000 documents. Afterwards, I fed the texts of the retrieved documents into the BERT classifier. Then, Birch internally divides into sentences the documents inserted into it to compute the new sentence similarity scores. Lastly, the output classification scores from the sentence representations were combined with the BM25+RM3 scores through linear interpolation.

Specifically, I used the following Equation 1 which aggregates all sentence-level scores during inference and finds the average of all sentence scores, which I consider the new document scores. Then, I integrated the old and new document scores by implementing the Formula 1 to acquire the end scores and my final initial document ranking.

$$Score_d = \lambda \cdot S_{doc} + (1 - \lambda) \cdot \sum_{i=1}^{n} w_i \cdot S_i \quad (1)$$

Where $\lambda$ is the linear interpolation parameter, $S_{doc}$ stands for old scores (BM25+RM3 scores), w is the weight parameter and $S_i$ are the new sentence scores. Finally, I tested my Birch setup with the 2017 NYT Common Core collection to ensure I received similar results to those reported by Zeynep Akkalyoncu [2].

### 3.3.2. The sentence-transformers (Sentence-BERT) test

I employed SBERT to compute my similarity scores which then were consumed by the two Equations (1, 2) that I explored. The w parameter of the mathematical Formula 1 was set to one. Afterwards, I tested the subsequent Equation 2 that uses the maximum sentence score from each document to compute the similarity score.

$$Score_d = \lambda \cdot S_{doc} + (1 - \lambda) \cdot max_{i=1}^{n} SBERT_i \quad (2)$$

Where $SBERT_i$ is the new maximum sentence score of each document. During my experimentation with SBERT, I found that a FSSV-MP (Fixed Size Sentence Vector with Mean Pooling – Formula 1) approach returned the best results. Lastly, the formula implementation of linearly interpolating the newly calculated SBERT (CLEF documents) similarity scores and my old scores (BM25+RM3) was motivated by Thakur et al. [34]

### 3.3.3. Fine-tuning the Language Model

In addition to BERT-Base which I used to acquire a basic understanding of the benefits of a DL model, I explored the PubMedBERT uncased model by Gu et al. [15]. I experimented with PubMedBERT, which is a domain specific word embeddings pre-trained model to help better capture the statistical relations of all the words in the corpus. Instead of training word embeddings, it was more feasible to use publicly available pre-trained embeddings. The reason I tested domain specific word embeddings in addition to BERT was that PubMedBERT embeddings have been reported to improve the training model's ability to generalise to a wide range of biomedical natural language tasks.

Also, this model is an extended pretraining version of the BERT-Base uncased model over a collection of PubMed abstracts. Gu et al [15] has largely described the vocabulary, architecture and the domain specific pre-training from scratch process. I trained PubMedBERT with a batch size of 8 documents during training, 4 documents during inference, and one epoch with 3, 6 and 12 evaluation steps. Furthermore, PubMed document length posed an issue which needed to be addressed by pre-processing the long text. For longer texts, these were cut-off and I used the first 512 tokens. The original BERT implementation truncates longer sequences automatically, but for PubMedBERT this option was not available.

### 3.3.4. Limitations

In the case of document classification with Birch I trained the proposed system with 3, 6 and 12 evaluation steps along with one epoch. The primary reason for this decision was the limited availability of computational resources and the extremely long execution time for only a single parameters setup. Furthermore, because I trained with an incomplete epoch, I consider the possibility that the reported results for B and C models in Table 1 in the Results section are not indicative of the highest retrieval performance which can be achieved by the system. I believe that the noticed performance gains can be increased further given the availability of enabling computational resources to set the epochs number as high as possible and terminate the training based on the error rates.

### 3.4. Measures

To evaluate the performance of all the models I used the CLEF-TAR evaluation scripts ([35], [36]). The measurements I used for appraising the baselines were AP (Average Precision), R (Recall), NCG@20 (Normalised Cumulative Gain after 20 documents) and the norm_area which is the area under the cumulative Recall curve normalised by the optimal area.

### 3.5. Statistical Comparison

To compare performance between conditions, I first used one-way ANOVAs (for which I am reporting p and F values) and then performed follow-up pairwise comparisons using T-tests if statistically significant results were found ($p < 0.05$). Bonferroni corrections were applied to determine which conditions were significantly different. For T-tests, I computed p-values and the effect size using Cohen's d values (please note that $0.5 < d < 0.8$ is considered a medium effect size, whereas $d > 0.8$ is a high effect size [37].). The results from the one-way ANOVA and pair-wise T-Tests procedures indicated statistical significance at $p < 0.05$ in many cases. The conclusions of my statistical tests are reported in Section 4. Finally, I report my evaluated metrics in Table 1. using the CLEF-TAR evaluation toolkit for my retrieved results.

## 4. RESULTS

### 4.1. Evaluating the Subset Retrieval application

In all the systems I investigated, when I applied the Subset Retrieval application I noticed important improvements in the average retrieval performance with both annual CLEF collections. Nonetheless, I had to implement this application as part of the CLEF eHealth TAR Task 2 in relation to the subset identification (as described in Section 1) from the potentially relevant studies. Therefore, the Retrieval Subset application was a positive precondition that I had to apply to all of my experiments.

### 4.2. Evaluating the proposed BERT based DL strategy

#### 4.2.1. Evaluating the Similarity Search with SBERT path

The average retrieval performance (in Table 1) I observed after performing the linear interpolation (G model) was better than the BM25+RM3 performance. Specifically, by interpolating with

SBERT I noticed a higher AP, norm_area and NCG@20 than the corresponding metrics I attained from the customised BM25+RM3 model with both datasets. Lastly, the average Recall recorded was 100%, but this is expected when evaluating these type of IR systems.

**4.2.2. Evaluating the Document Classification with BERT/PubMedBERT approach.**

The results in Table 1 I acquired using the Birch toolkit integrated with domain specific word embeddings (PubMedBERT) indicated no added benefit over the use of BERT-Base embeddings for the CLEF task at hand. In fact, I noticed the same retrieval effectiveness with both embeddings consumed by my document classification DL approach.

Table 1. Comparing the performance of the BM25+RM3 baseline model against the DL method variants I explored: Birch+BERT, Birch+PubMedBERT and SBERT+FSSV-MP and the corresponding interpolated model results. The term 'Normalised' refers to the application of Formula 1 when $\lambda = 0.8$ (from subsection 3.3.1.)

| Dataset | Model | AP | R | norm_area | NCG@20 |
|---|---|---|---|---|---|
| 2017 CLEF eHealth 30 SRs | BM25+RM3 (A) | $0.226^{D}$ | 1.0 | $0.778^{D}$ | $0.657^{D}$ |
| | Birch+BERT (B) | $0.213^{D}$ | 1.0 | $0.762^{G}$ | $0.539^{DG}$ |
| | Birch+PubMedBERT (C) | $0.213^{D}$ | 1.0 | $0.762^{G}$ | $0.539^{DG}$ |
| | SBERT+FSSV-MP (D) | $0.121^{ABCEFG}$ | 1.0 | $0.681^{AG}$ | $0.366^{ABCFGH}$ |
| | (B) + (C) Normalised (E) | $0.216^{D}$ | 1.0 | $0.76^{G}$ | $0.538^{DG}$ |
| | (B) + (D) Normalised (F) | $0.216^{D}$ | 1.0 | $0.76^{G}$ | $0.538^{DG}$ |
| | (B) + (E) Normalised (G) | $0.257^{D}$ | 1.0 | $0.858^{BCDEF}$ | $0.709^{BCEF}$ |
| 2018 CLEF eHealth 30 SRs | BM25+RM3 (A) | $0.202^{D}$ | 1.0 | $0.761^{G}$ | $0.587^{BCD}$ |
| | Birch+BERT (B) | $0.188^{G}$ | 1.0 | $0.701^{G}$ | $0.417^{AG}$ |
| | Birch+PubMedBERT (C) | $0.188^{G}$ | 1.0 | $0.701^{G}$ | $0.417^{AG}$ |
| | SBERT+FSSV-MP (D) | $0.111^{AEFG}$ | 1.0 | $0.778^{G}$ | $0.404^{AG}$ |
| | (B) + (C) Normalised (E) | $0.197^{D}$ | 1.0 | $0.757^{G}$ | $0.543^{G}$ |
| | (B) + (D) Normalised (F) | $0.197^{D}$ | 1.0 | $0.757^{G}$ | $0.543^{G}$ |
| | (B) + (E) Normalised (G) | $0.231^{BCD}$ | 1.0 | $0.902^{ABCDEF}$ | $0.749^{CDEF}$ |

What is more, the interpolated results from my document classification approach returned marginally better retrieval effectiveness than the BM25+RM3 model with both CLEF datasets. This modest increase in the retrieval effectiveness in comparison with the interpolated results from the SBERT method can be explained if we consider that I did not train the models with a high number of epochs.

**4.2.3. Identification of IR performance patterns**

To identify patterns, I focused on the G model from Table 1. because it returned the most promising results out of the DL variants I explored with both CLEF collections. I inferred from the following Figures 1 and 2 that the similarity search SBERT model with both 2017 and 2018 CLEF sets, it accomplished better AP, norm_area and NCG@20 performance for topics that include a smaller number of articles.

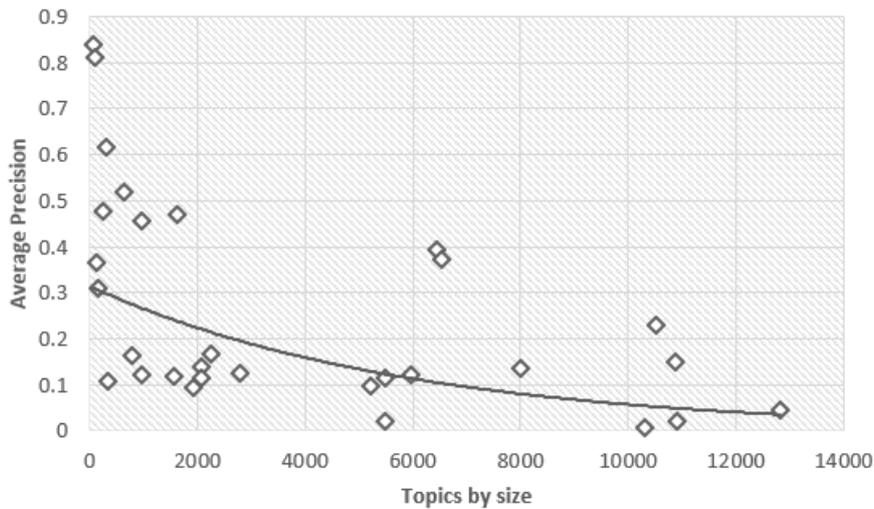

Figure 1. AP metric by the number of articles contained in the CLEF 2017 topics.

Statistical significance of the results. The ANOVA tests on the results from the 2017 and 2018 CLEF topics indicated significance. Specifically, for my experiments with the CLEF 2017 dataset I found the p-value = 0.00056 and the F-value = 4.613641. With the CLEF 2018 dataset I found the p-value < 0.00001 and the F-value = 9.975543. Suggesting that the samples are different. Later, the Bonferroni comparison post-hoc tests helped me identify which of the pairs of samples are significantly different from each other. I considered all pairs for simultaneous comparison in each SRs test collection and for each CLEF-TAR metric examined (as described in subsection 3.4) except for R since all models accomplished total Recall.

In Table 1., I connoted statistical significance between a pair of samples (models) by placing one model's identification letter to the upper right side of the other model's measurement result. The one-way ANOVA statistical test regarding the norm_area with the CLEF 2017 set returned a p-value equal to 0.0002 ($p < 0.05$) and the F statistic equal to 7.3152. The one-way ANOVA statistical tests related to the NCG@20 with the CLEF 2017 dataset returned a p-value of 0.0117 ($p < 0.05$) and the F statistic equal to 3.8310. In relation to the norm_area metric with the 2018 collection the F statistic was found to be 15.1147 and the p-value equal to 2.3417e-08 (i.e. $p < 0.05$). Lastly, for the NCG@20 metric with the 2018 dataset the one-way ANOVA F statistic was 12.05 and the p-value equal to 6.4033e-07 (i.e. $p < 0.05$).

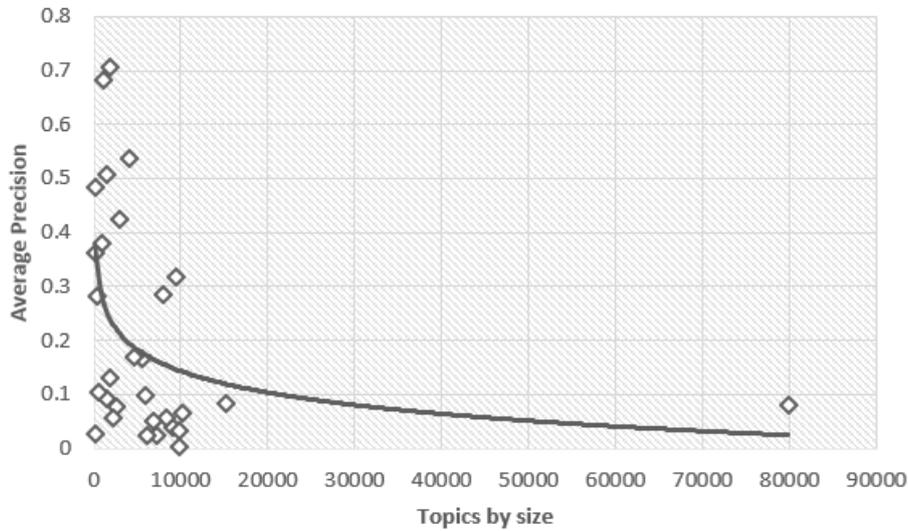

Figure 2. AP metric by the number of articles contained in the CLEF 2018 topics.

## 5. CONCLUSIONS

### 5.1. Summary

I focused on discovering how a DL strategy with two variants can impact the retrieval performance of healthcare SRs in the initial ranking for the CLEF eHealth TAR in Empirical Medicine Task. The first DL variation consisted of a document classification approach along with BERT and PubMedBERT embedding. The second DL variation was a similarity search path exploiting SBERT. I inferred from my document classification DL approach with the pretrained embeddings that it can contribute beneficially to the retrieval effectiveness of my IR baseline model for the TAR task examined. The findings from my SBERT experiments have driven me to the conclusion that the similarity search path with SBERT is a method that can improve the retrieval performance of the initial query against an already boosted BM25+RM3 (IR) model. Also, with the anticipated positive results from the correlation of the proposed DL strategy with my baseline IR model I add further evidence for the effectiveness of DL methods. Lastly, from the comparison of the two variants of the proposed DL strategy I concluded that the similarity search with SBERT approach is more effective in terms of retrieval performance.

### 5.2. Future Work

I aim to continue exploring different variations and new approaches to take full advantage of the many-mentioned BERT benefits. Specifically, my target is to improve the DL strategy by integrating alternative important embedding with Birch for instance FastText to evaluate the retrieval performance of the suggested embedding layer with the proposed system. Finally, I am looking forward to conduct a comparative study by combining a successful classification model such as Bidirectional LSTM with the recommended DL architecture.

# REFERENCES

[1] A. Tsertsvadze, Y.-F. Chen, D. Moher, P. Sutcliffe, and N. McCarthy, "How to conduct systematic reviews more expeditiously?" Syst. Rev., vol. 4, no. 1, p. 160, 2015. [Online]. Available: https://systematicreviewsjournal.biomedcentral.com/articles/10.1186/s13643-015-0147-7 [Accessed 2 Mar. 2021]

[2] J. Wallace, C. Byrne, and M. Clarke, "Improving the uptake of systematic reviews: a systematic review of intervention effectiveness and relevance" BMJ Open, vol. 4, no. 10, p. e005834, 2014. [Online]. Available: https://bmjopen.bmj.com/content/4/10/e005834.short [Accessed 2 Mar. 2021]

[3] A. C. Hardi and S. A. Fowler, "Evidence-based medicine and systematic review services at Becker Medical Library" Mo. Med., vol. 111, no. 5, pp. 416–418, 2014. [Online]. Available: https://www.ncbi.nlm.nih.gov/pmc/articles/PMC6172095/ [Accessed 2 Mar. 2021]

[4] L. Cappellato, N. Ferro, L. Goeuriot, and T. Mandl, "Ranking abstracts to identify relevant evidence for systematic reviews: The University of Sheffield's approach to CLEF eHealth 2017 Task 2: Working notes for CLEF 2017" in CEUR Workshop Proceedings, 2017, vol. 1866. [Online]. Available: https://core.ac.uk/download/pdf/153383184.pdf [Accessed 2 Mar. 2021]

[5] J. Chen et al., "ECNU at 2017 eHealth task 2: Technologically assisted reviews in empirical medicine" Ceur-ws.org. [Online]. Available: http://ceur-ws.org/Vol-1866/paper_99.pdf. [Accessed 2 Mar. 2021]

[6] J. Singh and L. Thomas, "IIIT-H at CLEF eHealth 2017 task 2: Technologically assisted reviews in empirical medicine" Ceur-ws.org. [Online]. Available: http://ceur-ws.org/Vol-1866/paper_116.pdf. [Accessed 2 Mar. 2021]

[7] C. Norman, M. Leeflang, and A. Neveol, "LIMSI@CLEF eHealth 2017 task 2: Logistic regression for automatic article ranking" Ceur-ws.org. [Online]. Available: http://ceur-ws.org/Vol-1866/paper_101.pdf. [Accessed 2 Mar. 2021]

[8] G. M. Di Nunzio, F. Beghini, F. Vezzani, and G. Henrot, "An interactive two-dimensional approach to query aspects rewriting in systematic reviews. IMS unipd at CLEF eHealth task 2" Unipd.it. [Online]. Available: http://www.dei.unipd.it/~vezzanif/papers/clef2017-1.pdf. [Accessed 2 Mar. 2021]

[9] G. E. Lee and A. Sun, "Seed-driven document ranking for systematic reviews in evidence-based medicine" in The 41st International ACM SIGIR Conference on Research & Development in Information Retrieval, 2018. [Online]. Available: https://dl.acm.org/doi/abs/10.1145/3209978.3209994 [Accessed 2 Mar. 2021]

[10] G. Eunkyung, "A study of convolutional neural networks for clinical document classification in systematic reviews: Sysreview at CLEF eHealth 2017" Edu.sg. [Online]. Available: https://dr.ntu.edu.sg/bitstream/10220/44958/1/A%20Study%20of%20Convolutional%20Neural%20Networks%20for.pdf. [Accessed 2 Mar. 2021]

[11] G. V. Cormack and M. R. Grossman, "Technology-assisted review in empirical medicine: Waterloo participation in CLEF eHealth 2017" Ceur-ws.org. [Online]. Available: http://ceur-ws.org/Vol-1866/paper_51.pdf. [Accessed 2 Mar. 2021]

[12] A. Carvallo and D. Parra, "Comparing word embeddings for document screening based on active learning" Ceur-ws.org. [Online]. Available: http://ceur-ws.org/Vol-2414/paper10.pdf. [Accessed 2 Mar. 2021]

[13] A. Carvallo, D. Parra, H. Lobel, and A. Soto, "Automatic document screening of medical literature using word and text embeddings in an active learning setting" Scientometrics, vol. 125, no. 3, pp. 3047–3084, 2020. [Online]. Available: https://link.springer.com/article/10.1007/s11192-020-03648-6 [Accessed 2 Mar. 2021]

[14]     J. Devlin, M.-W. Chang, K. Lee, and K. Toutanova, "BERT: Pre-training of deep bidirectional Transformers for language understanding" arXiv [cs.CL], 2018. [Online]. Available: https://arxiv.org/abs/1810.04805 [Accessed 2 Mar. 2021]

[15]     Y. Gu et al., "Domain-specific language model pretraining for biomedical natural language processing" arXiv [cs.CL], 2020. [Online]. Available: https://arxiv.org/abs/2007.15779 [Accessed 2 Mar. 2021]

[16]     A. Câmara and C. Hauff, "Diagnosing BERT with retrieval heuristics" in Lecture Notes in Computer Science, Cham: Springer International Publishing, 2020, pp. 605–618. [Online]. Available: https://link.springer.com/chapter/10.1007/978-3-030-45439-5_40 [Accessed 2 Mar. 2021]

[17]     A. Soleimani, C. Monz, and M. Worring, "BERT for evidence retrieval and claim verification" in Lecture Notes in Computer Science, Cham: Springer International Publishing, 2020, pp. 359–366. [Online]. Available: https://link.springer.com/chapter/10.1007/978-3-030-45442-5_45 [Accessed 2 Mar. 2021]

[18]     Z. Lu, P. Du, and J.-Y. Nie, "VGCN-BERT: Augmenting BERT with graph embedding for text classification" in Lecture Notes in Computer Science, Cham: Springer International Publishing, 2020, pp. 369–382. [Online]. Available: https://link.springer.com/chapter/10.1007/978-3-030-45439-5_25 [Accessed 2 Mar. 2021]

[19]     L. Sanchez, J. He, J. Manotumruksa, D. Albakour, M. Martinez, and A. Lipani, "Easing legal news monitoring with learning to rank and BERT" in Lecture Notes in Computer Science, Cham: Springer International Publishing, 2020, pp. 336–343. [Online]. Available: https://link.springer.com/chapter/10.1007/978-3-030-45442-5_42 [Accessed 2 Mar. 2021]

[20]     Z. Akkalyoncu Yilmaz, S. Wang, W. Yang, H. Zhang, and J. Lin, "Applying BERT to document retrieval with birch" in Proceedings of the 2019 Conference on Empirical Methods in Natural Language Processing and the 9th International Joint Conference on Natural Language Processing (EMNLP-IJCNLP): System Demonstrations, 2019, pp. 19–24. [Online]. Available: https://www.aclweb.org/anthology/D19-3004/ [Accessed 2 Mar. 2021]

[21]     P. Yang, H. Fang, and J. Lin, "Anserini: Enabling the use of Lucene for information retrieval research" in Proceedings of the 40th International ACM SIGIR Conference on Research and Development in Information Retrieval - SIGIR '17, 2017. [Online]. Available: https://peilin-yang.github.io/files/pub/Yang_etal_SIGIR2017.pdf [Accessed 2 Mar. 2021]

[22]     W. Yang, H. Zhang, and J. Lin, "Simple applications of BERT for ad hoc document retrieval" arXiv [cs.IR], 2019. [Online]. Available: https://arxiv.org/abs/1903.10972 [Accessed 2 Mar. 2021].

[23]     Roy, D., Ganguly, D., Bhatia, S., Bedathur, S. and Mitra, M. (2018). Using Word Embeddings for Information Retrieval. Proceedings of the 27th ACM International Conference on Information and Knowledge Management. [Online]. Available: https://dl.acm.org/doi/pdf/10.1145/3269206.3269277 [Accessed 2 Mar. 2021].

[24]     Palangi, H., Deng, L., Shen, Y., Gao, J., He, X., Chen, J., Song, X. and Ward, R. (2016). Deep Sentence Embedding Using Long Short-Term Memory Networks: Analysis and Application to Information Retrieval. IEEE/ACM Transactions on Audio, Speech, and Language Processing, 24(4), pp.694–707. [Online]. Available at: https://arxiv.org/pdf/1502.06922.pdf [Accessed 10 Apr. 2020].

[25]     N. Reimers and I. Gurevych, "Sentence-BERT: Sentence embeddings using Siamese BERT-networks" arXiv [cs.CL], 2019. [Online]. Available: https://arxiv.org/abs/1908.10084 [Accessed 2 Mar. 2021].

[26]     www.clef-initiative.eu. (n.d.). The CLEF Initiative (Conference and Labs of the Evaluation Forum) - Test Collections. [online] Available at: http://www.clef-initiative.eu/dataset/test-collection [Accessed 2 Mar. 2021].

[27]     GitHub. (2019). CLEFeHealth/CLEFeHealth2017IRtask. [online] Available at: https://github.com/CLEFeHealth/CLEFeHealth2017IRtask [Accessed 2 Mar. 2021].

[28] GitHub. (2019). CLEFeHealth/CLEFeHealth2018IRtask. [online] Available at: https://github.com/CLEFeHealth/CLEFeHealth2018IRtask [Accessed 2 Mar. 2021].

[29] Cochrane Training. (n.d.). training.cochrane.org. (n.d.). Ask the Expert. [online] Available at: https://training.cochrane.org/ask-expert [Accessed 2 Mar. 2021].

[30] Nih.gov. (2016). MEDLINE®/PubMed® Data Element (Field) Descriptions. [online] Available at: https://www.nlm.nih.gov/bsd/mms/medlineelements.html [Accessed 12 Oct. 2019].

[31] H. Imran and A. Sharan, "Thesaurus and Query Expansion" Psu.edu. [Online]. Available: https://citeseerx.ist.psu.edu/viewdoc/download?doi=10.1.1.212.9942&rep=rep1&type=pdf. [Accessed 2 Mar. 2021].

[32] N. Abdul-Jaleel, "UMass at TREC 2004: Novelty and HARD" Umass.edu. [Online]. Available: https://scholarworks.umass.edu/cgi/viewcontent.cgi?article=1185&context=cs_faculty_pubs. [Accessed 2 Mar. 2021].

[33] GitHub. (n.d.). castorini/anserini. [online] Available at: https://github.com/castorini/anserini/blob/master/src/main/java/io/anserini/search/SearchArgs.java [Accessed 2 Mar. 2021].

[34] N. Thakur, N. Reimers, J. Daxenberger, and I. Gurevych, "Augmented SBERT: Data augmentation method for improving bi-encoders for pairwise sentence scoring tasks" arXiv [cs.CL], 2020. [Online]. Available: https://www.researchgate.net/publication/344734344_Augmented_SBERT_Data_Augmentation_Method_for_Improving_Bi-Encoders_for_Pairwise_Sentence_Scoring_Tasks [Accessed 2 Mar. 2021].

[35] E. Kanoulas, D. Li, L. Azzopardi, and R. Spijker, "CLEF 2017 Technologically Assisted Reviews in Empirical Medicine Overview" CEUR Workshop Proc., vol. 1866, p. 29, 2017. [Online]. Available: https://pureportal.strath.ac.uk/en/publications/clef-2017-technologically-assisted-reviews-in-empirical-medicine- [Accessed 2 Mar. 2021].

[36] E. Kanoulas, D. Li, L. Azzopardi, and R. Spijker, "Clef 2018 Technologically Assisted Reviews in Empirical Medicine Overview" CEUR Workshop Proc., vol. 2125, Jul 2018. [Online]. Available: https://pureportal.strath.ac.uk/en/publications/clef-2018-technologically-assisted-reviews-in-empirical-medicine- [Accessed 2 Mar. 2021].

[37] S.Cohen, " Perceived stress in a probability sample of the united states." 1988. Apa.org. [Online]. Available: https://psycnet.apa.org/record/1988-98838-002. [Accessed 2 Mar. 2021].